# Interworking Scheme Using Optimized SIP Mobility for MultiHomed Mobile Nodes in Wireless Heterogeneous Networks

Paolo Dini, Jaume Nin-Guerrero, Josep Mangues-Bafalluy

IP Technologies Area
Centre Tecnològic de Telecomunicacions de Catalunya (CTTC)
Av. Canal Olimpic s/n – 08860 Castelldefels, Barcelona, Spain
{paolo.dini; jaume.nin; josep.mangues}@cttc.es

Lillian Dai, Sateesh Addepalli

Cisco Research
425 East Tasman Drive
San Jose , CA 95134 - United States
{lildai, sateeshk}@cisco.com

**Abstract — Nowadays, mobile users wish to use their multi-interface mobile devices to access the Internet through network points of attachment (PoA) based on heterogeneous wireless technologies. They also wish to seamlessly change the PoAs during their ongoing sessions to improve service quality and/or reduce monetary cost. If appropriately handled, multihomed mobile nodes offer a potential solution to this issue. In this sense, the management of multihomed mobile nodes in heterogeneous environment is a key research topic. In this paper, we present an improvement of SIP mobility (pre-call plus mid-call mobility) to support seamless mobility of multihomed mobile nodes in heterogeneous wireless networks. Pre-call mobility is extended to associate user identifier (i.e. SIP URI) and interface identifiers (i.e. IP addresses). The multiple addresses of a mobile device are weighted by the user to create a priority list in the SIP server so as to guarantee resilient reachability of mobile nodes and to avoid unnecessary signaling through wireless links, thus saving radio resources. Then, three variations of mid-call mobility, called hard, hybrid and soft procedures, are also proposed. Their main aim is to minimize, or even avoid, packet losses during interface switching at the mobile node. The proposed solutions have been implemented in a wireless heterogeneous testbed composed of 802.11 WLAN plus 3.5 cellular network, which are fully controlled and configurable. The testbed has been used to study the performance and the robustness of the three proposed mid-call mobility procedures.**

## I. INTRODUCTION

Nowadays, mobile devices are equipped with multiple wireless interfaces to access Internet services through multiple access technologies. Terminal mobility support across such a heterogeneous environment must be provided so that a user may change his/her network point of attachment without losing his/her connectivity and maintaining a high service quality.

One solution to this is to have loosely-coupled networks and multihomed devices in such networks [1]. A multihomed mobile node is a device with at least one valid address assigned to each of its multiple interfaces and able to simultaneously use these addresses.

The framework to manage multihoming may be based on different protocols. IETF MEXT group is working on a modification of MIPv6 to address this issue [2]. Other protocols may also be used for the same purpose, namely SCTP [3], HIP [4] or SIP [5].

The present paper improves SIP pre-call and mid-call mobility procedures ([6]) to manage multihomed mobile nodes. Reachability of the device is assured when it is changing its IP addresses and smooth quality degradation is guaranteed when ongoing sessions switch from one interface of the device to another, i.e. during vertical handoff.

The choice of SIP as protocol to manage multihomed mobile nodes is due to different reasons. The user of a SIP client running on a multihomed mobile node is identified by a unique SIP URI, such as alice@proxy.com. The URI is registered in its SIP server (i.e. proxy.com in this example) together with its multiple IP addresses, through which it is reachable by correspondent nodes (CNs). Then, SIP provides a natural way to associate user identifiers (i.e. SIP URI) and interface identifiers (i.e. IP addresses) without the need for defining complex procedures, like the registration of multiple care-of-addresses for a home address to create multiple binding cache entries for MIPv6 [7]. Apart from that, SIP has been originally designed to manage multimedia sessions and has been selected as the session management signaling protocol for the Internet Multimedia Subsystem (IMS) architecture [5]. Additionally, it does not require any modifications to existing network layer protocols, which eases deployment. It also works with both IPv4 and IPv6. It has low implementation complexity, even in a heterogeneous wireless environment, since it is a text-based protocol. However, the use of SIP may limit the provision of data services (i.e. TCP traffic) that are not supported by SIP signaling so far.

The main contributions of this paper are:

- A pre-call mobility optimization procedure that associates SIP URI and IP addresses of the device interfaces by means of a prioritized list defined by the user, thus guaranteeing resilient reachability of the multihomed mobile nodes and, at the same time, avoiding unnecessary signaling during interface switching;
- Two new mid-call mobility optimization procedures that extend the hard procedure presented in our previous work [8], namely hybrid and soft, whose main aim is to minimize (or even, reset) the number of lost packets during interface switching. An optimization of the hard procedure in [8] is also presented in this paper.

The proposed procedures are experimentally evaluated using the EXTREME Testbed® [9]. In particular we use MUSA [10], the UMTS-HSDPA architecture of the testbed, to have complete access and control on network configuration parameters (e.g. UMTS bearers settings, number of users in the cell, coverage area).

The paper is organized as follows. Section II introduces the related work on the topic. Section III defines the three interface switching procedures (hard, hybrid and soft) by showing the signaling messages. In Section IV, the testbed used for the experiments is introduced and the analysis of the achieved results is reported. Conclusions and future work are finally presented in Section V.

## II. RELATED WORK

[6] reports some instances of SIP signaling to support terminal mobility. These examples have been considered in the recent literature to provide terminal mobility over wireless heterogeneous networks. Some examples are in the following.

In [11] the authors propose a modification of standard SIP signaling and architecture to provide support for vertical handoffs without disruption of real-time multimedia services. They introduce two new entities in the standard architecture to handle the proposed signaling scheme. They also implement the considered scenario in a testbed. The solution seems valid though no experimental result is shown in the paper. Furthermore it is not totally compliant with standard SIP architecture by [6].

In [12] the authors use SIP to perform vertical handoffs and also to evaluate its performance when switching between WLAN and UMTS. Mobile nodes are not multihomed in this work. Then, network specific registration and IP assignment procedures are necessary during the handoff (e.g. DHCP and PDP Context Activation procedures). The analysis concludes that the handoff from WLAN to UMTS is very slow because of the PDP Context Activation procedure in UMTS, so seamlessly interface switching is not feasible due to packet losses. Then, they propose a soft-handoff based on the duplication of data traffic during the switching time: the MN transmits simultaneously though both the interfaces to decrease the number of lost packets. The application at the receiver is in charge of discarding the arriving duplicate packets. This proposal does not take into account the benefits of having multihomed nodes. A double use of the bandwidth occurs during the interface switching due to the duplicate packets transmission, thus leading to an inefficient use of the already scarce radio resources.

In [13] the authors introduce the concept of soft handoff on a SIP based vertical handoff in a pure IP wireless network to ensure no packet loss during the interface switching. In particular, a new SIP procedure between the old and the new serving base station is added to the standard SIP mid-call mobility procedure based on the insertion of JOIN header option in the re-INVITE message. Consequently, they define a new architecture with the presence of a SIP user agent also in the base stations. The paper presents a sort of distributed mobility scheme where different network elements share the control of the handoff. Nevertheless, it is a modification of the standard SIP architecture defined in the [6].

In [14] the authors present the issues related to SIP mobility without giving a solution. The concept of registering the multiple interfaces of the mobile node in the SIP server is introduced in the paper but no explanation on how SIP can manage the multihomed hosts.

A step forward towards SIP-managed multihoming is in [15]. The authors propose an association scheme between the SIP URI of a user and its multiple IP addresses (each associated to one network interface) managed by the SIP server. The interfaces are sorted by signal strength and traffic load within the SIP server. According to this scheme, the paper is describing only pre-call mobility. But no explanation about the signaling for mid-call mobility is provided.

The work proposed in the paper has the following contributions with respect to the studied literature:

- use of legacy SIP mobility procedures to manage multihomed mobile nodes and their mobility in wireless heterogeneous networks;
- optimization of SIP mobility procedures to obtain:
  - resilient mobile node reachability;
  - preservation of radio resources by limiting the signaling and unnecessary data traffic through wireless links;
  - no packet loss during the interface switching;
- experimental evaluation of the optimized SIP mobility procedures using real equipment and fully configurable wireless heterogeneous networks.

## III. SIP MOBILITY OPTIMIZATION FOR MULTIHOMED MOBILE NODES

SIP mobility is divided into two phase pre-call and mid-call mobility. Pre-call mobility preserves the reachability of a device for incoming requests when it moves among IP networks. The process involves the re-registration of a SIP client within its SIP server when it moves from a network to another. Mid-call mobility maintains ongoing sessions when a device switches its IP address. In this case, the mobile node (MN) re-invites the CN to a session with its new adopted IP address. Then, both nodes stop the voice communication using the old address and restart it using the new one with the new parameters [6].

The optimization introduced in this paper count with the procedures explained in the following.

### *A. Pre-call mobility*

The main aim of our pre-call mobility optimization is to avoid sending registration update messages to register the new interface IP address during interface switching of the active sessions, thus saving radio resources as well as speeding up the mid-call mobility procedure and giving at the same time a resilient reachability to mobile nodes.

In compliance with SIP registration procedure, the MN registers its SIP URI and all its available interfaces within its SIP server by sending a REGISTER message. In our scheme the MN has to add weights associated to each of its interfaces in the *q* field of the *contact header*. The weights are defined on user preference to create a priority list of the registered interfaces within the SIP server. The priority list is used by the SIP server to forward signaling messages to MN.

For instance, a user of a multihomed MN with a WLAN and cellular interface can prefer to be always reachable, and then can assign higher priority to its cellular interface due to the higher coverage of the network. Therefore, cellular interface is used to forward signaling messages first. WLAN is used for signaling only when no answer is received from the higher priority interface, so as to have always a resilient connection between the mobile node and its SIP server.

At the end of this procedure, all the active IP addresses of the MN are registered and listed, based on user priority, in the SIP server, that does not need of any registration update message during the mid-call mobility.

Summarizing, registration update procedure is performed:

- after the network specific registration and IP address assignment procedures, when the MN is switched on,
- when a new interface is available at the MN; for example, when it enters in a coverage area of a new network and obtains a new IP address,
- when the user decides to change his/her list of interface priority.

Note that this priority list is used only to forward signaling messages and does not apply to data traffic. The choice of the interface to be used for data packet is taken when the session starts, i.e. at the moment of the dispatch/reception of the INVITE message by the MN, based on output of a decision algorithm. The interface can also change during the duration of the session, for example, because MN detects that another of its interface can provide a better session quality. This situation is managed by mid-call mobility.

### B. *Mid-call mobility*

Three proposed optimizations of the mid-call mobility are proposed in the following. They are called hard, hybrid and soft procedure. Actually, the first is an extension of our previous work in [8], and the last two are new proposals whose main aim is to minimize the number of lost packets during the interface switching.

As stated in the previous section, no REGISTER message is sent to the SIP server during the mid-call procedure due to the previous registration of all the available interfaces and the relevant IP addresses.

#### 1) *Hard Procedure*

When the interface switching is triggered, the MN sends a re-INVITE message and subsequently data packets through the new interface. The old interface is closed to forward data traffic, so as to realize a break-before-make procedure. The CN continues sending its data traffic to the MN old interface till the reception of the re-INVITE message and the dispatch of the OK message.

#### 2) *Hybrid Procedure*

Similarly to the previous case, when the interface switching is triggered, the MN sends a re-INVITE message and subsequently data packets through the new interface. The old interface is kept open. CN continues sending its data packets through the old interface till the reception of the re-INVITE message and the dispatch of the OK message. Then, it changes the destination IP address. The MN closes the old interface at the reception of the OK message from the CN.

#### 3) *Soft Procedure*

When the interface switching is triggered, the MN sends a re-INVITE message through the new interface but differently from the previous cases, data packets still travel through the old interface that is kept active. CN continues sending its data packet through the old interface till the reception of the re-INVITE message and the dispatch of the OK message. Then, it changes the destination IP address. The MN closes its old interface after the reception of the OK message from the CN.
A make-before-break procedure is realized in such a case.

For the sake of clarity and comprehension, Figure 1 summarizes the three proposed solutions.

## IV. EXPERIMENTAL ANALYSIS

The present section aims at experimentally analyzing the three procedures of mid-call mobility introduced in this paper when applied to manage the interface switching for multihomed MNs.

Experimental tests have been carried out using three nodes of the EXTREME Testbed® [9]. One PC is acting as multi-interface mobile node, another as WLAN access point and SIP server and the third as correspondent node. All the machines run a Fedora Linux OS with kernel 2.6.17.11. Mobile node is equipped with a WLAN card (Atheros chipset) and a UMTS-HSDPA card (OPTION GT-MAX). MUSA [10] has been used as UMTS-HSDPA network that permits a complete control on cellular network configuration parameters.

The experiments consist in performing a VoIP call and changing the interface during the duration of the call. The transitions from WLAN to UMTS-HSDPA and from UMTS-HSDPA to WLAN are analyzed. Each call lasts for one minute and the switching is executed in the middle of the call.

Packet losses, one way delay and perceived voice quality are calculated to assess the degradation of the voice session during the interface switching in the three introduced procedures and to understand its causes. The extended E-Model ([16], [17]) is chosen to estimate the perceived voice quality, since this method is assumed as more accurate than E-Model [18] for the bursty packet losses occurring in our test scenario. A sliding window has been used for these calculations: packet delay and losses collected during a certain time window (60msec) are averaged; these averaged values are used to calculate the R-factor for that time window. The metrics are all calculated at IP level by parsing captured packet traces. Therefore, the effect of playout buffer is not considered to calculate the R-factor. However, IP level delay obtained in our tests is always below 10 msec for WLAN and between 40 and 80 msec for UMTS-HSDPA, so the playout buffer effect is negligible in the calculation of the R-factor and then on the proposed analysis. It is worth noting that results are averaged over 50 repetitions of the same experiment before plotting the considered graph.

Three VoIP codecs are studied, namely G711, G729, G723.1. Table 1 summarizes the main characteristics of the studied codecs.

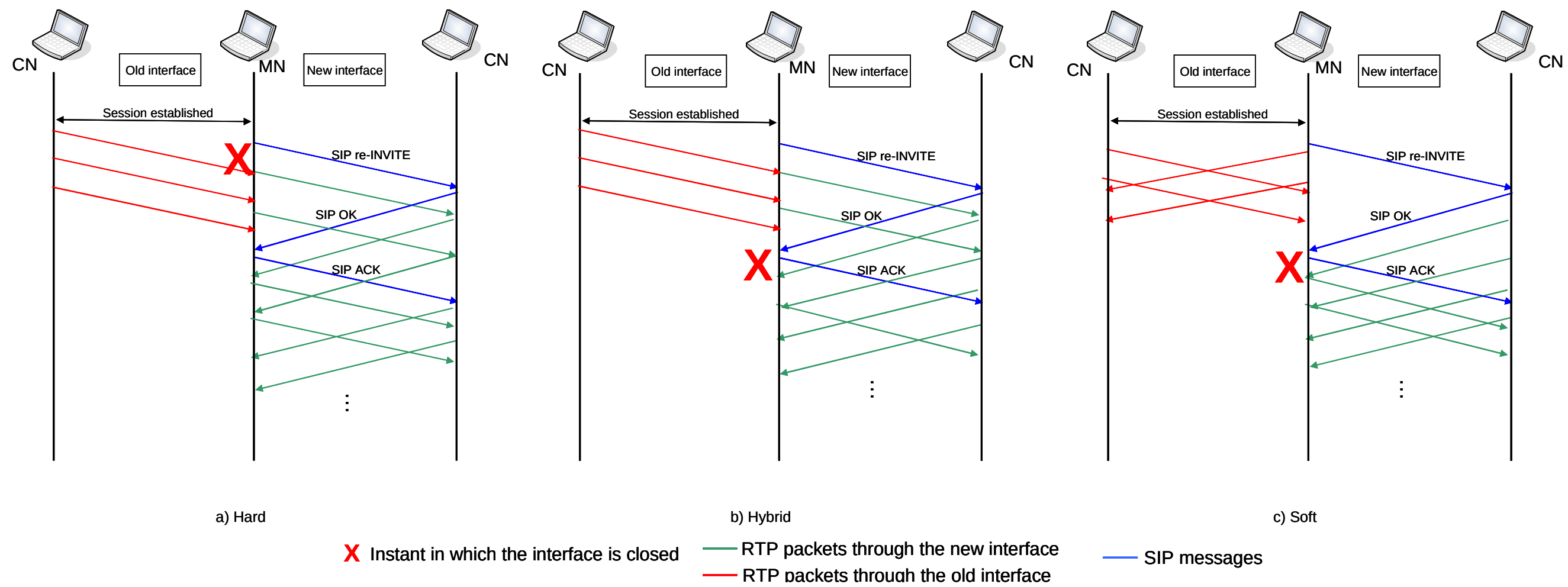

Figure 1. Message charts of the three interface switching procedures (from left to right): hard, hybrid and soft.

**TABLE 1: Used Voice Codecs and their packet level characteristics**

| Codec | Bitrate (kbps) | Packet inter-arrival time (msec) | RTP Packet size (bytes) |
|---|---|---|---|
| G711 | 64 | 20 | 160 |
| G729 | 8 | 20 | 20 |
| G723.1 | 6.3 | 30 | 24 |

We modified *Ekiga* open source application [19] to act as VoIP traffic generator and SIP client supporting the three interface switching procedures. SIP Express Router (SER) [20] software is used as SIP server. Multihoming is handled by means of *iproute2* Linux utility suite [21] using IP source based routing.

Experimental tests are divided into two measurement campaigns. In the first campaign, we discuss the main characteristics of the introduced procedures and their performance when applied to reduce voice quality degradation during mid-call mobility. The second campaign aims at argue the robustness of the introduced procedures when taking into account realistic network conditions (i.e. network congestion).

### A. *Performance Analysis*

Figures 2 and 3 present the results of the measurement campaign when the switching is performed from UMTS-HSDPA to WLAN and from WLAN to UMTS-HSDPA, respectively. Each graph represents the behavior of the R-factor of one codec during the call when the interface switching procedures are applied. Uplink (UL) and downlink (DL) are separated in the graphs so as to analyze both directions in detail.

In-lab networks are underloaded in this campaign, then we can notice a satisfactory value of R-factor (R>70) in the stationary conditions, i.e. before and after the change of the interface. In this testing scenario, WLAN can provide better voice quality than UMTS-HSDPA due to its higher bitrate. The same characterization applies for UMTS-HSDPA due to the different bitrate in UL and DL. Moreover, the three codecs present different values of the R-factor due to their intrinsic characteristics.

It is interesting to note that data packets from CN to MN (i.e. DL) start passing through the new interface in distinct instants according to the applied interface switching procedure. In particular, the soft procedure is the last in changing the interface as can be foreseen looking at the relevant message chart on Figure 1.

It is also worth highlighting that one way delay observed in every performed test is very low (around 10 msec for WLAN and between 40 and 80 msec for UMTS-HSDPA) and no spike is observed during the transition from one interface to the other due to the design of the three proposed procedures. Consequently, packet loss, rather than delay, is the metric with more impact on the perceived voice quality according to [16] and [17].

Let us consider the transition from one interface to the other in Figure 2. UL does not present any problem for all the introduced procedures. A transitory is needed to reach the new stable situation with a new value of the R-factor depending on the new network and the codec. On the contrary, DL presents a breakdown (equal to about 0,5) before reaching the stationary value of the R-factor, when considering the hard procedure. Breakdown is due to a high number of lost packets during the transition in the hard procedure (see Figure 4-left, which shows the percentage of lost packets during the interface switching). On the other hand, no packet is lost when the hybrid and soft procedures are considered (see Figure 4-left), thus justifying the absence of a breakdown when these other two procedures are applied. The breakdown due to packet losses when the hard procedure is applied cannot be properly appreciated in Figure 3, due to the lower value of R-factor in UMTS-HSDPA. Anyway, Figure 4-right confirms that even in this case several data packets are lost. It is interesting to notice that the number of lost packets in the hard procedure, when performing the transition from WLAN to UMTS-HSDPA is higher than from UMTS-HSDPA to WLAN. WLAN has a higher bitrate than UMTS-HSDPA; then a higher number of data packets can travel, and be lost, during the INVITE-OK handshake.

Such results are confirmed by the audio tests in the lab: a listener at CN can hear a voice glitch when hard procedure is applied. On the contrary, no interruption is appreciated in the case of hybrid and soft procedures.

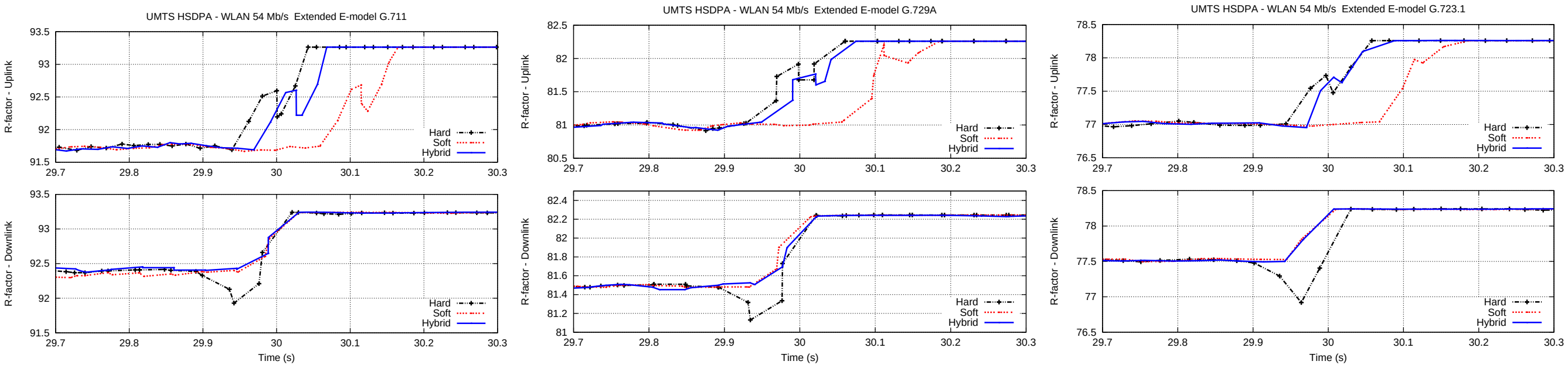

Figure 2. R-factor calculated for the transition from UMTS-HSDPA to WLAN interface of the G711, G729, G723.1 codecs (left to right) and for each of the considered procedure

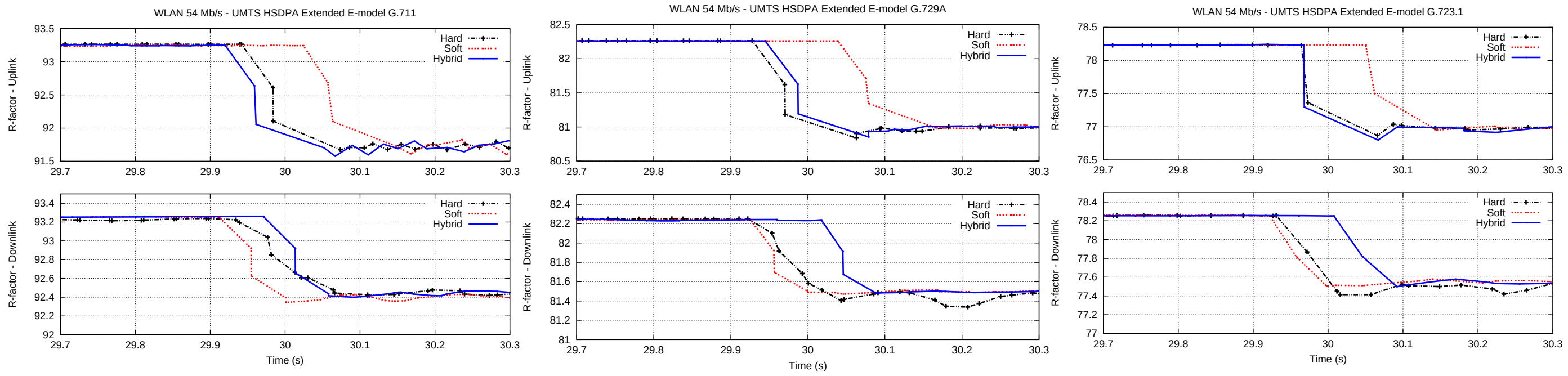

Figure 3. R-factor calculated for the transition from WLAN to UMTS-HSDPA interface for each of the G711, G729, G723.1 codecs (left to right) and for each of the considered procedure

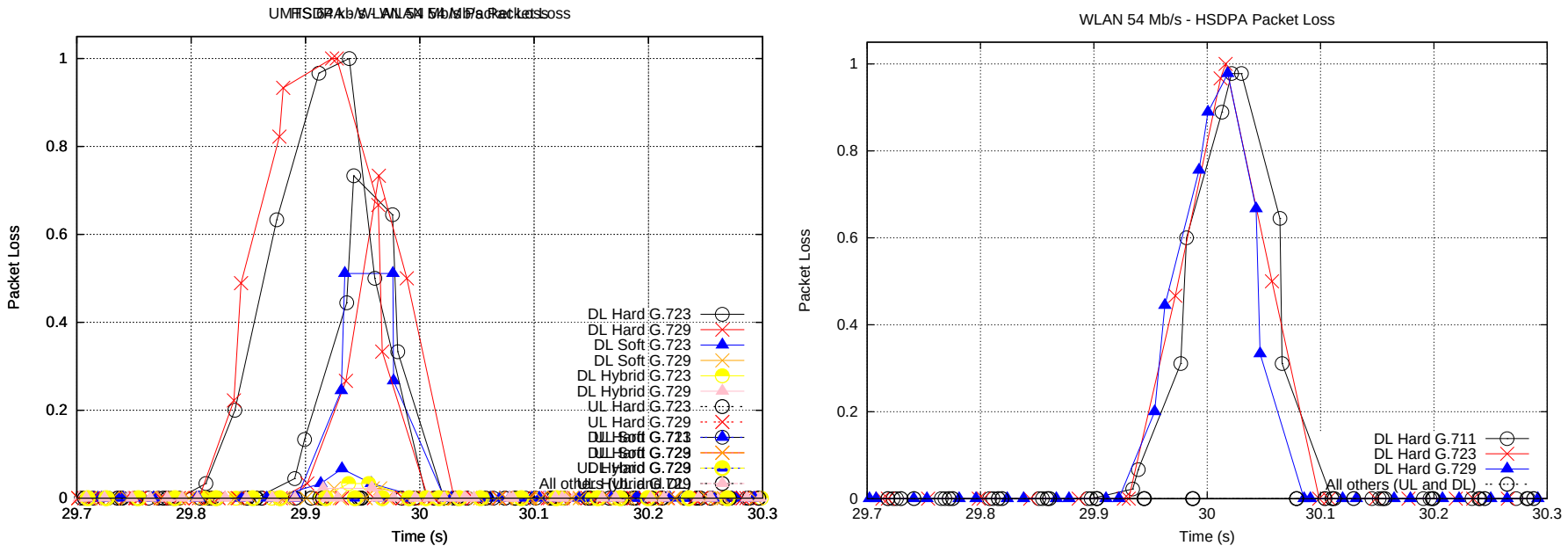

Figure 4. Percentage of packet losses for the transition from UMTS-HSDPA to WLAN (left) and from WLAN to UMTS-HSDPA (right) for each of the considered codec and for each of the considered procedure

The problem of the hard procedure is intrinsic in the “break-before-make” nature of the procedure, i.e. to close the old interface before the last packet to the old interface IP address reaches the MN. Shutting down the old interface after the completion of the re-INVITE-OK handshake procedure minimizes the probability of loosing packets sent by the CN to the old interface IP address.

Though the three considered codecs implement packet loss concealment, they reveal some weakness to bursty packet losses, i.e. the breakdown commented above. Then, soft or hybrid procedures are appropriate solutions to handle seamless interface switching for voice services.

## *B. Robustness*

The main aim of this sub-section is to study the robustness of the three procedures, to understand the limits of the proposed solutions in a close-to-real scenario.

The first case to analyze is when re-INVITE (or OK) message is lost, for example due to channel errors. The SIP client has to re-send the message. The hard procedure is the most affected by the time wasted during the re-INVITE-OK handshake, as stated in the previous sub-section. Its performance will be still worse. The other two procedures (hybrid and soft) are more robust, since old interface data socket remains open till the OK message arrival and no packet will be lost even in this case.

In real conditions, the decision to switch to another interface has to be taken when the new network can provide a better quality to the active sessions, e.g. the new network has a higher available bitrate than the old. If the two networks present very unbalanced bitrate, then the procedures can present some drawbacks. Actually, hard procedure is not particularly affected by this situation; it breaks the link before the instauration of the new connection and then data packets are lost anyway till the end of the re-INVITE-OK handshake. On the other hand, even the soft and hybrid procedure can present packet losses, if the re-INVITE-OK handshake time is faster than the arrival of the last packet sent by the CN to the old interface IP address.

The in-lab wireless heterogeneous network has been properly set to emulate the scenario presented above. The switching is performed from a very low bitrate to a high bitrate network. In particular, the transition from UMTS at 64 kbps to WLAN at 54 Mbps is analyzed.

In Figure 5 the percentage of packet losses of the G729 and G723.1 codecs are reported when the three interface switching procedures are applied. G711 has not been tested in this case due to its higher bandwidth requirements. Figure 5 shows that very few packets are lost when soft and hybrid procedures are applied. In particular, the hybrid presents a bit lower number of lost packets because CN changes destination IP address just after the reception of the re-INVITE, whereas the soft waits until the completion of the re-INVITE-OK handshake.

The results confirm the robustness of the hybrid and the soft procedures (more than the hard), since very few packets are lost also considering a constraining scenario.

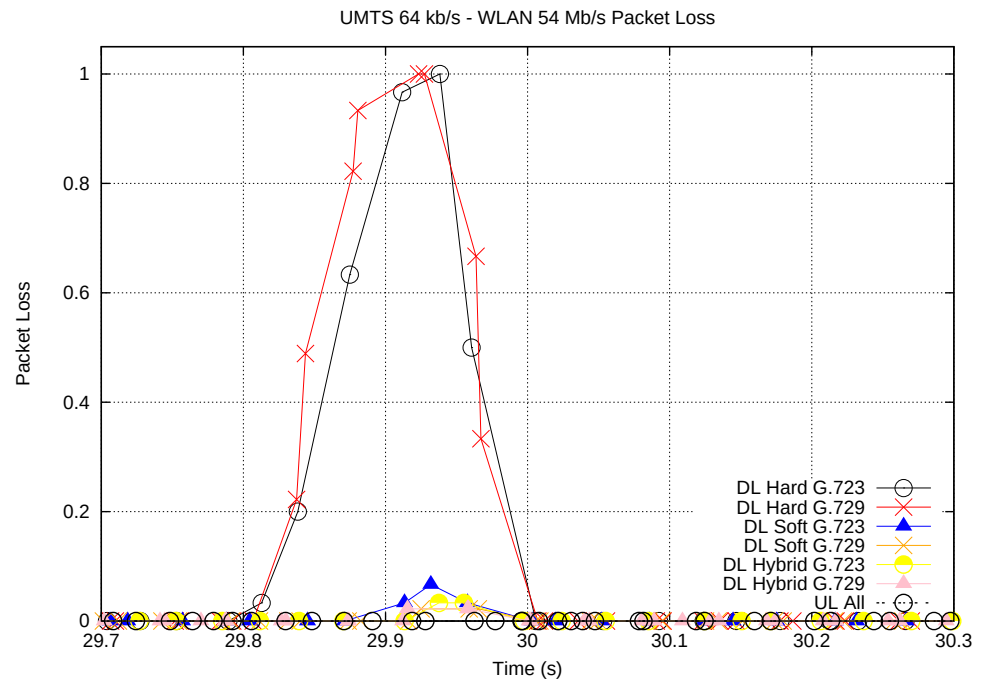

Figure 5. Percentage of packet losses for the transition from UMTS at 64 kbps and WLAN at 54 Mbps for G729 and G723.1 codecs and for each of the considered procedure

## Conclusions and Future Work

In this paper, we propose an optimization of SIP mobility (i.e. pre-call and mid-call) to support interface switching for multihomed mobile nodes in heterogeneous wireless networks. We define an extension of pre-call mobility in multi-addresses scenario through a weighted association between user identifier (i.e. SIP URI) and interface identifiers (i.e. IP addresses). The procedure guarantees resilient reachability of mobile nodes and avoids unnecessary signaling through wireless links, thus saving radio resources.

We also propose three variations of mid-call mobility, by introducing the so-called hard, hybrid and soft procedures. Experimental evaluation is carried out in a fully controlled wireless heterogeneous testbed and considering VoIP traffic. Results prove that the three considered VoIP codecs (G711, G729, G723.1) are not robust to the bursty packet losses occurring in our test scenarios, though implementing packet loss concealment. Experimental trials also prove that soft and hybrid procedures minimize packet loss, even in a constraining scenario. Then, the two introduced procedures are good schemes to manage seamless interface switching for VoIP services.

The main limitation of our SIP-based approach is that it is currently restricted to UDP-based connections. The main goal of our future work is to extend SIP signaling to also support TCP connections, i.e. data services.

## Acknowledgments

This project has been made possible through joint collaboration with Cisco Advaced Architecture & Research Group and by Spanish Ministry of Science and Innovation under grant TEC2008-06826/TEC (project ARTICO).